# Interplay of substrate inclination and wettability on droplet impact dynamics

**Nilamani Sahoo[1], Gargi Khurana[1], A R Harikrishnan[2], Devranjan Samanta[1, a, #] and Purbarun Dhar[1, b, #]**

[1]Department of Mechanical Engineering, Indian Institute of Technology Ropar, Rupnagar–140001, India

[2] Department of Mechanical Engineering, Indian Institute of Technology Madras, Chennai–600036, India

[#] Corresponding authors:
[a] E-mail: devranjan.samanta@iitrpr.ac.in
[b] E–mail: purbarun@iitrpr.ac.in
[b] Phone: +91-1881-24-2173

## Abstract

Experimental investigations were carried out to elucidate the role of surface wettability and inclination on the post-impact dynamics of droplets. Maximum spreading diameter and spreading time were found to decrease with increasing inclination angle and normal Weber number ($We_n$) for superhydrophobic (SH) surfaces. The experiments on SH surfaces were found to be in excellent agreement with an existing analytical model, incorporated with the modifications for the oblique impact conditions. Energy ratios and elongation factor were also measured for different inclination angles. On inclined SH surfaces, different features like arrest of secondary droplet formation, reduced pinch-off at the contact line and inclination dependent elongation mechanism were observed. Contrary to SH surfaces, hydrophilic surfaces show opposite trends of maximum spreading factor and spreading time with inclination angle and normal Weber number respectively. This was due to the dominance of tangential kinetic energy over adhesion energy and gravitational potential at higher inclination angles. Finally, colloidal solutions of nanoparticles were used to elucidate slip and disjoining pressure on SH and hydrophilic surfaces, respectively. Overall, the article provides a comprehensive picture of post- impact dynamics of droplets on inclined surfaces encompassing a broad spectrum of governing parameters like Reynolds number (Re), Weber number (We), degree of inclination and surface wettability.

# 1. INTRODUCTION

Dynamics of droplets upon impact on solid surfaces has been an area of tremendous interest in research and development since the pioneering work by Worthington[1]. Understanding droplet dynamics is vital for applications like fuel injection, inkjet printing, spray painting, pesticides and fumigation, electronic component welding, ice formation on aircrafts[2, 3], etc. Drop impact dynamics on surfaces is a complex phenomenon due to interplay of surface physics[4, 5], interfacial chemistry and fluid mechanics[6, 7]. Post-impact drop dynamics is driven by non-dimensional numbers like Weber number (We), Reynolds number (Re), Capillary number (Ca) as well as the wettability of the surface[5].

Previous studies of drop impact on hydrophobic or superhydrophobic surfaces (SH) have shown various modes like splash, deposition or rebound[2, 8-10]. Studies have reported the maximum spreading diameter[4, 11-14], temporal spreading and recoil dynamics of the drops[15-16]. Most analytical models[11, 12, 16-19] based on energy conservation principle were focused on prediction of the maximum spreading factor along the surface. Clanet et al.[4] and Bejan and Gobin[17] predicted the maximum spreading ratio based on scaling analysis. In some studies[11, 18] the dynamic contact angle and its temporal variations were also considered to estimate the spreading.

Along with the maximum spreading diameter, the spreading time is another important parameter. Richards et al.[20] measured the capillary-inertial time for impacted droplet onto SH surfaces by regression analysis of experimental data. Bartolo et al.[21] studied the dewetting phenomena of droplet impact onto hydrophobic surfaces and calculated the relaxation time for capillary-viscous regime. In addition, many researchers considered the advective time scale, to study the impact outcomes of droplet onto the surfaces[22-25]. Wang et al.[24] and Lin et al.[25] conducted experiments on horizontal surfaces to find spreading time using universal scaling law.

Compared to normal impact cases, drop impact on inclined planes is a relatively lesser studied area[26-29]. Sikalo et al.[30] reported asymmetry in the spreading factor i.e. the differences in the deformation at the front and back of the droplet after impact. Chiarot et al.[31] and Zheng et al.[32] studied grazing impact of high velocity continuous drop streams on inclined SH surfaces and observed that the shape and structure of the rebounding stream is influenced by the frequency of the drop ejection and velocity. Yeong et al.[33] investigated the dependence of drop dynamics on Weber number. Antonini et al.[34] distinguished six distinct impact regimes at wide range of We for drop impact on tilted hydrophobic and SH substrates. LeClear et al.[35] observed the transition from the superhydrophobic Cassie–Baxter regime to the fully-wetted Wenzel regime while studying the impact of water drops on inclined textured SH surfaces. Aboud and Kietzig[36] observed the oblique splashing threshold on surfaces of different wettability at high impact velocities.

The present study aims to provide a better understanding of droplet impact on oblique surfaces and explore the roles of governing non dimensional parameters like We, Re as well as impact of surface inclination and wettability on impact dynamics. Experiments have been performed with various test fluids and impact conditions to encompass a vast spectrum of governing We and Re numbers. Surface conditions were varied to study the effect of hydrophilicity and superhydrophobicity at different inclinations. Subsequently the experimental data was matched with an exisitng mathematical model[37, 38] of droplet spreading dynamics. Existing analytical models from the previous studies along with the selected one were initially compared with the present experimental results for normal impact. The selected analytical model was shown to be in good agreement with SH cases.

## 2. MATERIALS AND METHODOLOGIES

The experimental setup (Fig. 1) consisted of a digitally controlled drop dispenser (Holmarc Opto-Mechantronics Pvt. Ltd., India) mechanism that discharged constant volume drops from a 100 μl glass syringe. Images were recorded using a high speed camera (Photron FASTCAM SA4) mounted with a G-type AF-S Macro lens of focal length 105mm (Nikkor, Nikon). The camera was operated at 1024 x 1024 pixels resolution at 3600 fps for imaging. Experiments were performed at ambient conditions (25°C) on hydrophilic and SH surfaces. For hydrophilic surfaces, sterile glass slides were thoroughly cleaned with acetone and DI water and then dried in hot air oven. The SH surfaces were created on similar glass substrates using superhydrophobic spray coating (Ultra Tech International Inc., USA). DI water was used as the primary fluid for droplets. $SiO_2$ water nanocolloids (2.5 and 5 wt. %) was used to understand the influence of fluid viscosity on impact dynamics of droplets without changing the surface tension. $SiO_2$ nanoparticles (average size 7 nm) are hydrophilic and disperse very well in water forming a stable colloidal phase. The aqueous solution of $SiO_2$ behaves like a single component homogeneous fluid rather than as a heterogeneous dispersion.

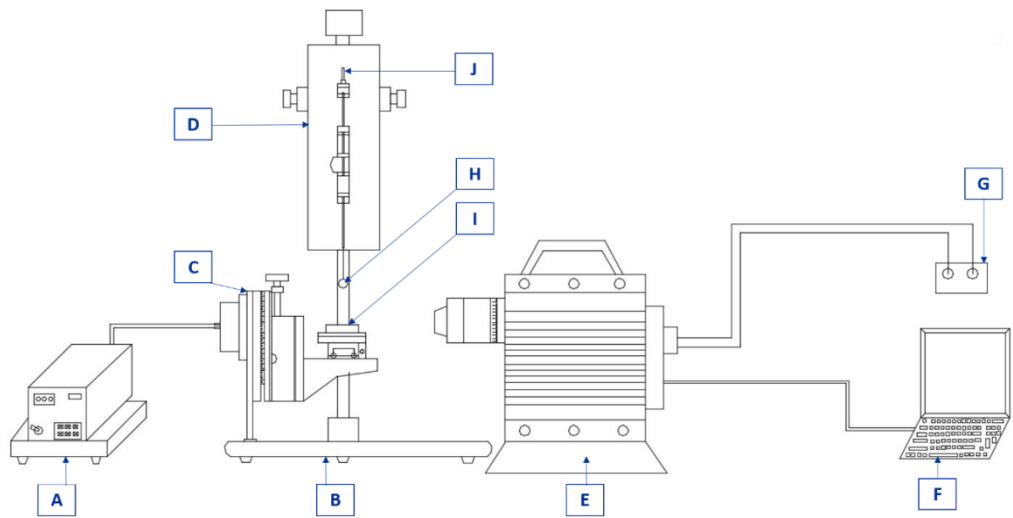

**Fig. 1** Schematic of the experimental setup (A) Drop dispenser controller (B) Base (C) Substrate inclination apparatus with backlight arrangement (D) Syringe pump (E) High speed camera (F) Laptop (G) Power source (H) Drop (I) Target surface (J) Syringe.

Conversely, to study the effect of surface tension without change in the fluid viscosity, different concentrations of sodium dodecyl sulfate (SDS) dissolved in DI water (0.25 and 0.50 % of the Critical Micelle Concentration (CMC)) were used as test fluids. The static contact angles of these fluids on the hydrophilic and SH substrates were measured using a contact angle goniometer (Holmarc Opto-Mechantronics Ltd., India). The static contact angle for water was observed to be nearly 137°. The nature of contact angle hysteresis on such surfaces has been reported in literature by the present authors[39]. To quantify the exact drop diameter from the dispensed volume, several experiments were performed for different test fluids using the mechanical dispenser unit and a precision electronic balance (Shimadzu ATX, Japan). The target substrates were provided with the requisite inclination using an inclination mechanism with angle control accurate to ± 0.5°. The properties of various test fluids are tabulated in Table 1. Viscosity and surface tension of test fluids were measured using a Rheometer (MCR 102, Anton Paar, Germany) and pendent drop analysis, respectively. The variation in properties was within ± 5 %.

**Table 1:** Properties of the test liquids at 25 °C

| Liquid | $\rho_l$ (kg/m$^3$) | $\sigma_l$ (N/m) | μ (mPas) |
|---|---|---|---|
| DI water | 997 | 0.072 | 0.89 |
| Silica nano-colloid (2.5 wt %) | 1015.8 | 0.072 | 4.0 |
| Silica nano-colloid (5 wt %) | 1032.2 | 0.072 | 54.0 |
| Surfactant solution (0.25 CMC) | 997 | 0.054 | 0.89 |
| Surfactant solution (0.5 CMC) | 997 | 0.045 | 0.89 |

The plane of impact was inclined at Φ with respect to horizontal plane. Accordingly, the droplet velocity $v_i$ was composed of normal ($v_i\cos\Phi$) and tangential ($v_i\sin\Phi$) components at the plane. The impact velocity was calculated using the theory of free falling objects with the assumption of negligible air drag. It was ensured that the needle discharged the drop vertically and that there was no cross stream flow of ambient air during the droplet flight. The initial diameter of the droplet was used as the characteristic length scale for calculation of governing non-dimensional parameters. The Weber number in normal direction of the impact plane $We_n = \rho v_i^2 (\cos\Phi)^2 d_i / \sigma$ was defined based on the normal velocity component $v_i\cos\Phi$, density (ρ), diameter ($d_i$) and fluid surface tension (σ) (subscripts i : initial condition before impact; n : normal direction). $We_n$ was varied between 2.5 and 127 ($We_n$ = 2.5–89 for water, $We_n$ = 3–127 for surfactant solutions (0.25 and 0.5 wt % CMC), and $We_n$ = 2.5–89 for aqueous solutions of silica particles (2.5 and 5 wt %)). Similarly, the Reynolds number based on normal velocity is defined as $Re_n = \rho d_i v_i \cos\Phi/\mu$. $Re_n$ was varied from 13.5 to 4587 ($Re_n$=812-4587 for water, $Re_n$=13.5-1084 for silica colloids and $Re_n$=705.5-4587 for surfactant solutions). Non-dimensional parameters in tangential directions were calculated using the tangential velocity component ($v_i \sin\Phi$). The initial droplet diameters for DI water, 0.25 and 0.5 wt. % CMC were ~ 2.9, 2.71, and 2.52 mm, respectively. The initial droplet diameters for 2.5 and 5 wt. % silica colloids are almost equal to the initial diameters for the water droplets.

## 3. MATHEMATICAL FORMULATION

In comparison to different theoretical and semi-analytical models [4,7,11,12] the present model proposed by Yonemoto and Kunugi (to be referred to as Y-K hereafter) [37,38] seems to match better with our experimental findings on SH surfaces (fig. 4a) Based on conservation of energy principle, Y-K derived the analytical expression for maximum spreading diameter ($\Psi_m$), expressible as

$$We = 3*\left[\frac{27}{16}\frac{r_i^2}{l_m^2}\left(\frac{We}{Re}\right)\psi_m + (1-\cos\theta)\psi_m^2 - \left(\frac{l_m}{r_i}\psi_m \sin\theta\right) - \left(\frac{2}{3}\frac{l_m \rho g r_i}{\sigma}\right) - 4 + \left(\frac{S_d}{\pi r_i^2}\right)\right] \quad (1)$$

Where, $r_i$ and $l_m$ are droplet radius before impact and the maximum spreading thickness after drop impact on the surface respectively and $\theta$ is the static contact angle for the particular substrate. Density and surface tension of the fluid are denoted as $\rho$ and $\sigma$ respectively. Acceleration due to gravity is denoted as g and $S_d$ is the deformed surface area of post impact drops. In Eq. (1), the Re and We are only defined for horizontal surfaces. For the case of impact on inclined planes, We and Re was replaced by the earlier defined $We_n$ and $Re_n$. The deformed surface area for SH and hydrophilic surfaces have been denoted as $S_{d,SHS}$ and $S_{HPS}$ respectively. In accordance with the experimental observations of the post impact events, analytical expressions of $S_{d,SHS}$ and $S_{HPS}$ are deduced as follows:

**Case 1: Estimation of $S_{d,SHS}$**

The $S_{d,SHS}$ is defined as the harmonic mean of the spherical cap ($S_{cap}$) case at low We and flattened disc surface ($S_{disc}$) at high We (Fig. 2) and expressed as :

$$S_{d,SHS} = \frac{2 S_{cap} S_{disc}}{S_{cap} + S_{disc}} \tag{2}$$

$$S_{cap} = \pi \left( r_m^2 + l_m^2 \right) \tag{3}$$

$$S_{disc} = \pi \left( r_m^2 \right) + 2\pi r_m l_m \tag{4}$$

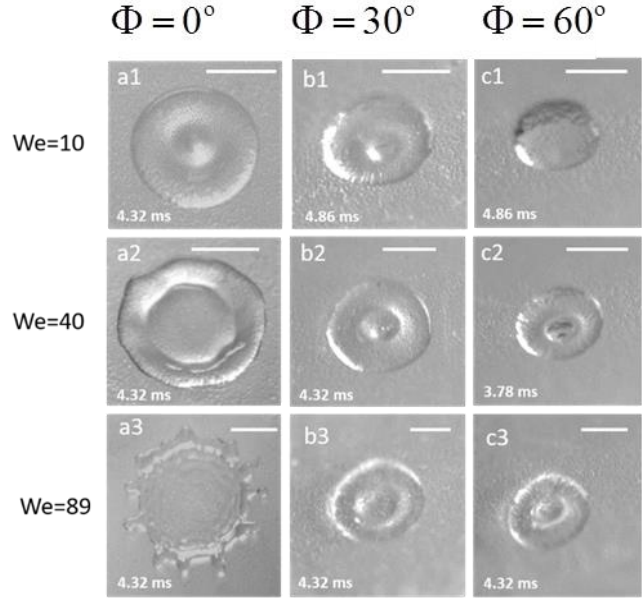

**Fig.2.** Top view of the droplets in their maximum spread state after impact on SH surfaces at different We and surface inclination. The top view of the droplets after impact is similar to flat discs. The scale bars in all cases correspond to 2.9 mm.

### Case 2: Estimation of $S_{HPS}$, $\Phi < 30°$

The spreading mechanism along an oblique hydrophilic surface is observed to be strikingly different from that of oblique SH surfaces. The droplets approach a flattened disc phase after impacting onto the hydrophilic surfaces at lower We ($We_n = 2.5-10$) (Fig. 3). Therefore, the moment at which maximum spreading diameter is attained by the droplet, the deformed surface area can be approximated as a flattened disc, expressible as

$$S_{d,HPS} = 2\pi r_m l_m \tag{5}$$

For a drop impact study, Yonemoto and Kunugi[38] considered the average of the dynamic contact angles at the moment of maximum spreading diameter. This approach showed better agreement with experimental data. Therefore, the contact angle has been proposed to be considered as the average of the static and dynamic contact angles (Eqn. 6) for the present analytical model. When the droplets achieve the maximum spreading radius after impact, the dynamic contact angle is considered. Again, the static contact angle ($\theta_s$) is considered for SH surface as reported in previous study[40].

$$\theta_{av} = \frac{\theta_s + \theta_d}{2} \qquad (6)$$

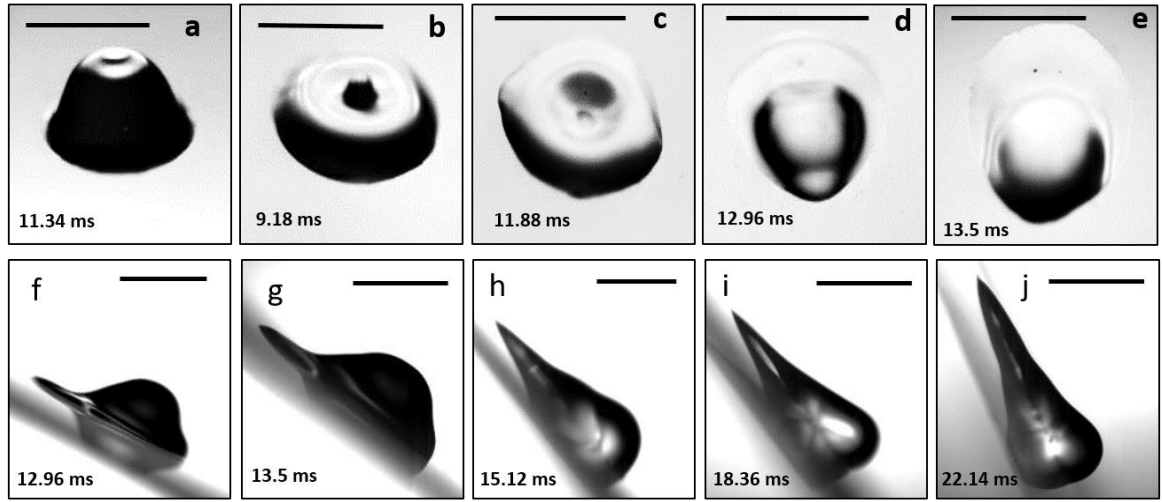

**Fig. 3.** Top views of the droplet at the respective maximum spreading diameter for inclined hydrophilic surfaces at (a) 0° (b) 10° (c) 20° (d) 30° and (e) 40° at We=10. The camera was positioned vertically with respect to the target surface to image post impact dynamics on the glass substrate. The motion in each case is from the top to the bottom. Front view images at maximum spreading for (f) 30° (g) 40° (h) 50° (i) 60° (j) 70°. The test liquid is water and the impact velocity is 0.5 m/s for each case. The scale bars in all cases correspond to 2.9 mm.

### Case 3: Φ>30° for hydrophilic surface

At different inclinations, the drops attain different shapes, especially upon impact on hydrophilic surfaces. After impact, the drops change from oval to cusp shape with elongated conical structure along the inclined plane[41]. In the present experiments, similar structures were observed to slide down at higher surface inclinations i.e. Φ>30° (Fig. 3). This phenomenon was observed at lower impact velocities (We=10). Therefore, Eqn. (5) overestimates the deformed surface area as the maximum spreading diameter increases with increasing substrate inclination. Since there is no cap formation, the deformation energy ($E_d=S_d\sigma_l$) in the modified analytical model is neglected (Eqn. 6) to approximate the conditions for $\Phi >30^o$ as $\quad S_{d,HPS}=0 \quad\quad\quad (7)$

The present study also explores the effects of surface inclination and fluid properties on the spreading time. For droplets, the spreading time is generally normalized with three characteristic time scales, viz. advection time, capillary-inertia time and capillary-viscous time. The spreading time is dependent on factors like surface wettability, impact velocity and liquid properties. To understand the role of surface inclination on spreading time ($\tau_s$), a universal scaling correlation of spreading time reported in literature has been considered[25]. Instead of using Weber number (We), Weber number in normal direction ($We_n$) was used in the estimation of the spreading time as $\quad \dfrac{\tau_s}{\left(\dfrac{\rho r_m^3}{\sigma}\right)^{0.5}}=0.92 We_n^{-0.43} \quad\quad (8)$

## 4. RESULTS AND DISCUSSION

The experimental data and its validation with the modified analytical model is categorized into three parts: (a) Superhydrophobic (SH) surface, (b) hydrophilic (HP) surface and (c) nano colloids as a special test fluid to reveal the role of slip on SH surface and disjoining pressure on hydrophilic surface.

Before starting our discussion on different parameters related to droplet impact, figure 4 highlights the droplet dynamics after impact on SH and hydrophilic surfaces both inclined at $60^o$ with respect to horizontal plane. In case of SH surfaces, droplets are sliding as well as deforming tangentially along the surface. Sliding absorbs substantial portion of the tangential K.E. and resists deformation along the inclined surface. After attaining maximum spreading, droplets bounce orthogonally to the inclined surface. In case of HP surfaces sliding is absent. Rather the drop pins at a certain location and undergoes subsequent elongation in tangential direction. Contrary to SH surface there was no orthogonal bouncing off the surface after attaining maximum spreading.

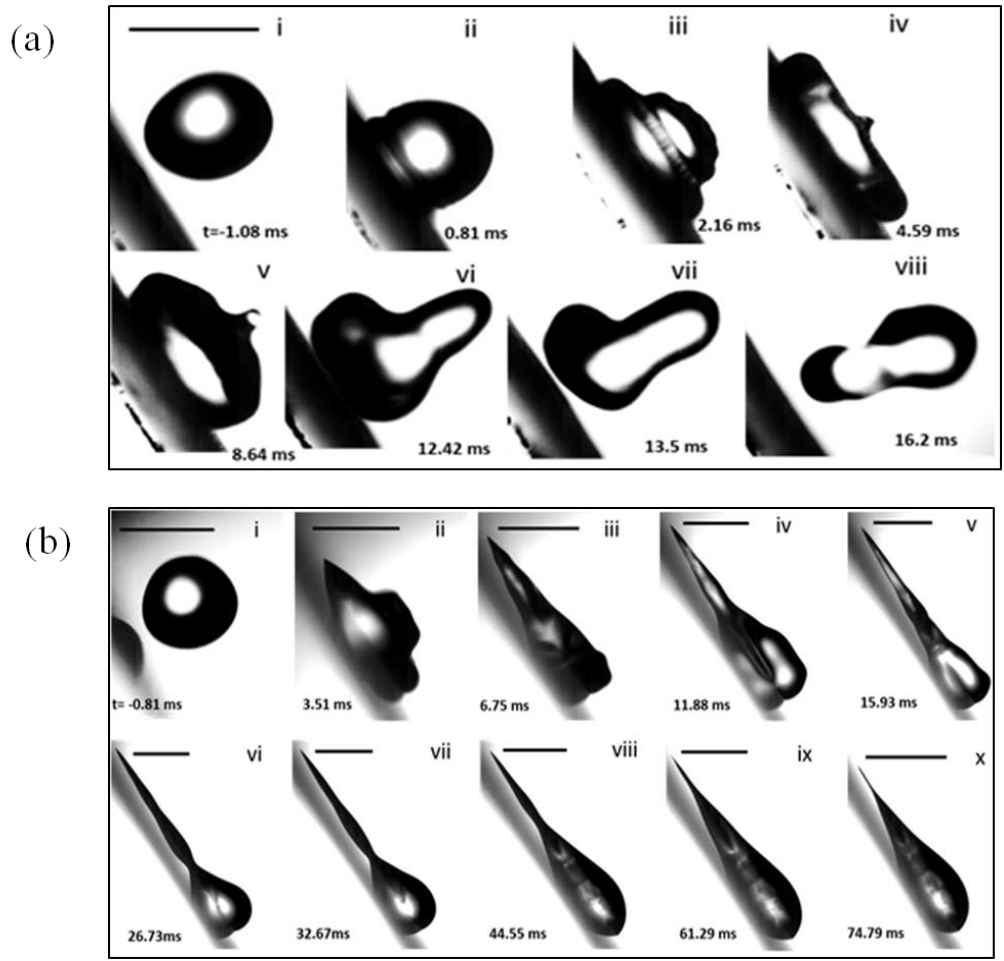

**Fig.4** Droplet dynamics after impact at 60° inclination (a) on SH surface and (b) HP surface. Moment of impact is taken as t=0. Hence in both (a-i) and (b-i) time with '–' sign signifies time before impact. Scale bar of all figures represent 2.9 mm.

### a. <u>Superhydrophobic surface (SH)</u>
### I. Maximum spreading factor ($\Psi_m$) and spreading time ($\tau_s$) on SH surfaces

Droplet dynamics after impact can be characterized by maximum spreading factor ($\Psi_m$) and spreading time ($\tau_s$). Fig.5 (a) compares the present experimental maximum spreading factor ($\Psi_m$) on horizontal SH surfaces for normal impact cases with different analytical and semi-empirical models [5, 7, 11, 12, 39, 42] including the present modified model. Three different fluids (water, surfactants and silica nanocolloids) were used. The static contact angles for the fluids are provided given in Table S1 (supplementary information). From fig. 5a it is evident that for wide spectrum of impact velocity (We=10-89), the Y-K model is in better agreement with the experimental data in comparison to other models. Consideration of the role of interfacial and deformation energy components in Y-K model leads to better matching than other models.

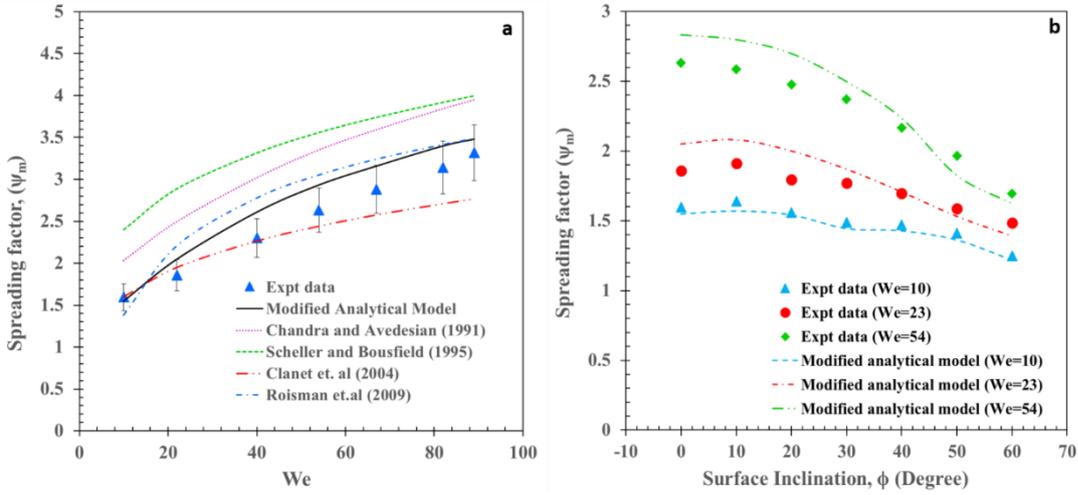

**Fig.5** (a) Spreading factor vs. Weber number is plotted to validate of experimental results of normal impact cases on SH surface with the Y-K analytical model. (b) Variation of maximum spreading factor for DI water on SH surface at different inclinations and impact velocities [$Re_n$=812-3807 (Re=1624-3807) and $We_n$=2.5-54 (We=10-54)].

Equation (1) was modified to replace Re and We with $Re_n$ and $We_n$ in normal direction to the inclined surface for comparison of oblique impact cases. Fig. 5b shows that the modified Y-K model is able to predict the general decreasing trend of the experimental observations at different impact velocities. For a given initial velocity (fixed We), the maximum spreading diameter decreases with increasing angle of inclination (decreasing $We_n$). Again for a fixed Φ, spreading factor increases with the increase of $We_n$. With the increase in inclination, sliding of the whole droplet predominates over the radial spreading, thereby leading to reduced spreading factor.

In order to understand the impact of viscosity on spreading dynamics, experiments were also conducted with fluids of different viscosity (water and aqueous colloid of silica nanoparticles (2.5 and 5 wt. %)). Fig.6 (a) and (b) illustrate the variation of maximum spreading factor with $Re_n$ for water and 2.5 wt % of silica nanocolloid respectively. From these figures it can be observed that the maximum spreading factor is highest for the horizontal surface i.e. Φ =0$^o$ case and reduces with increasing inclination angles. Further, for a given inclination angle, the spreading factor is observed to increase with $Re_n$. Comparison of figure 6(a) and (b) shows that for the range of Re covered in the present experiments, the spreading factor seems to be unaffected due to change in viscosity. Since the trend of spreading factor is mainly governed by deceleration of radial expansion flow and faster sliding motion along the tilted plane with increase in Φ, the change in fluid viscosity does not influence its behaviour.

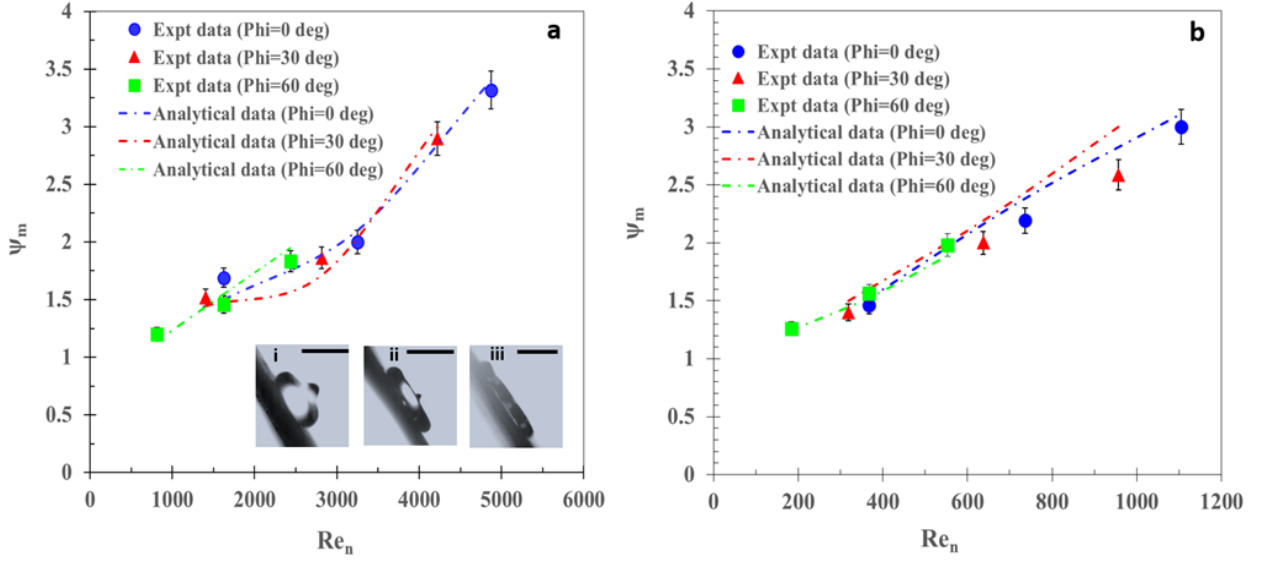

**Fig.6** (a) Effect of viscosity on maximum spreading factor $\Psi_m$ for water droplets on SH surface. Spreading time vs. $Re_n$ is plotted Inset: (i), (ii) and (iii) show the time evolution of droplet spreading dynamics along the inclined plane at Φ =60°. The magnitude of scale bar is equal to 2.9 mm for each figure. (b) The maximum spreading factor of aqueous colloid of silica droplets (2.5 wt. %) on SH surface to showcase the effects of fluid viscosity on the impact dynamics.

In addition to the maximum spreading factor, the maximum spreading time $\tau_m^* = \dfrac{\tau_s}{\left(\dfrac{\rho r_m^3}{\sigma}\right)^{0.5}}$, normalized by the capillary inertial time ($t_{capillary-inertia} = \sqrt{\rho r_i^3/\sigma}$) has also been calculated. Fig. 7 shows the normalized spreading time based on maximum droplet radius. From figure 7 a-d it is evident that the non-dimensional maximum spreading time decreases monotonically with increase in the normal Weber number $We_n$. As seen from Fig 7, for a fixed We, at Φ =0° the droplets achieve maximum spreading leading to the maximum spreading time also. The radial expansion flow is enhanced by the initial kinetic energy, which in turn promotes the fluid to attain maximum spreading time at Φ =0°.

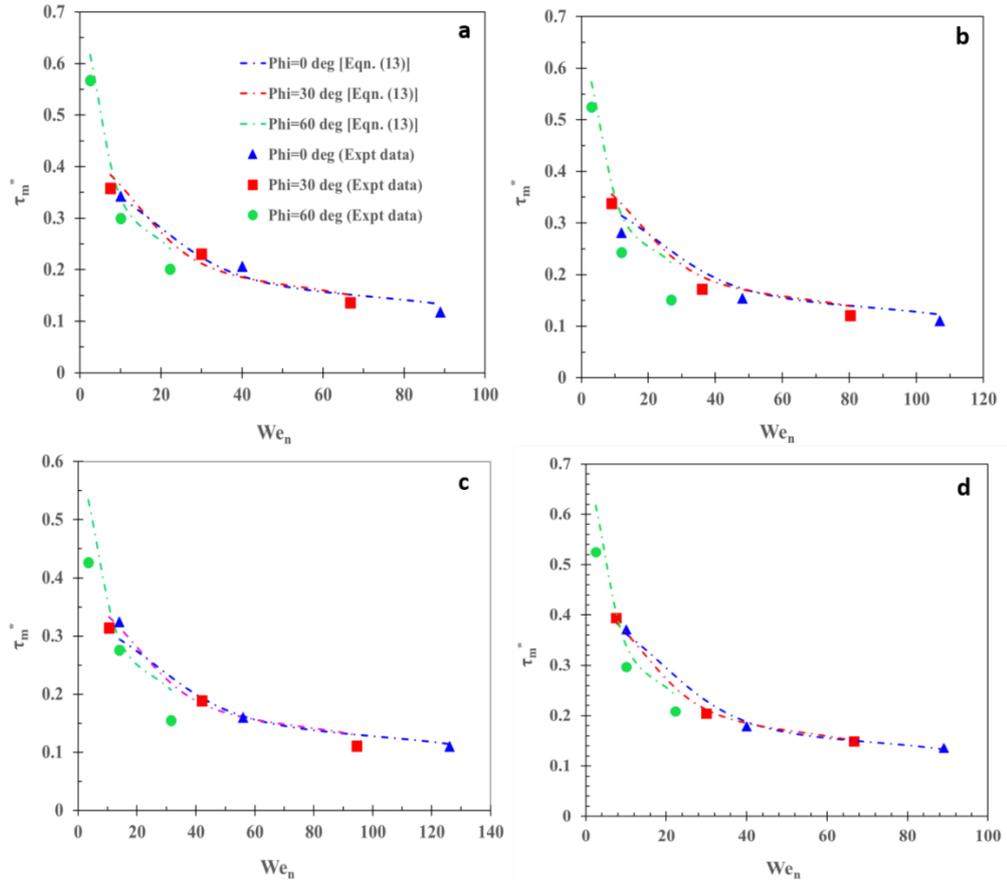

**Fig.7** (a) Variation of maximum spreading time normalized with capillary-inertial time considering maximum drop radius for water, and similarly (b), (c) and (d) for 0.25 and 0.5 CMC SDS solution and 2.5 wt % aqueous solution of silica nano-colloid respectively on SH surface. Symbols in Fig 5(b)-(d) are same as in Fig.5 (a).

## II. Temporal evolution of elongation and spreading factors on SH surfaces

Fig.8 (a) illustrates the variation of elongation factor β and spreading factor Ψ with respect to the non-dimensional time ($t^*=tv_i/d_i$) on SH horizontal surfaces (at $\Phi = 0^o$). The elongation factor ($\beta=h/d_i$) has been defined as the ratio of the vertical height of the drop at the onset of recoiling to the initial drop diameter (left hand ordinate). It can be readily observed from Fig.8 (a), that the spreading dynamics of liquid drops impacting onto oblique SH surfaces is strikingly different from the recoiling mechanism instigating the elongation of liquid drops along normal direction. The recoiling phenomenon begins when the drop achieves its maximum spreading diameter. Therefore, the elongation factor follows a different trend compared to the spreading factor, where the bouncing drop propagates into the air medium eventually generating secondary drops (inset (iii) of Fig.8 (a)).

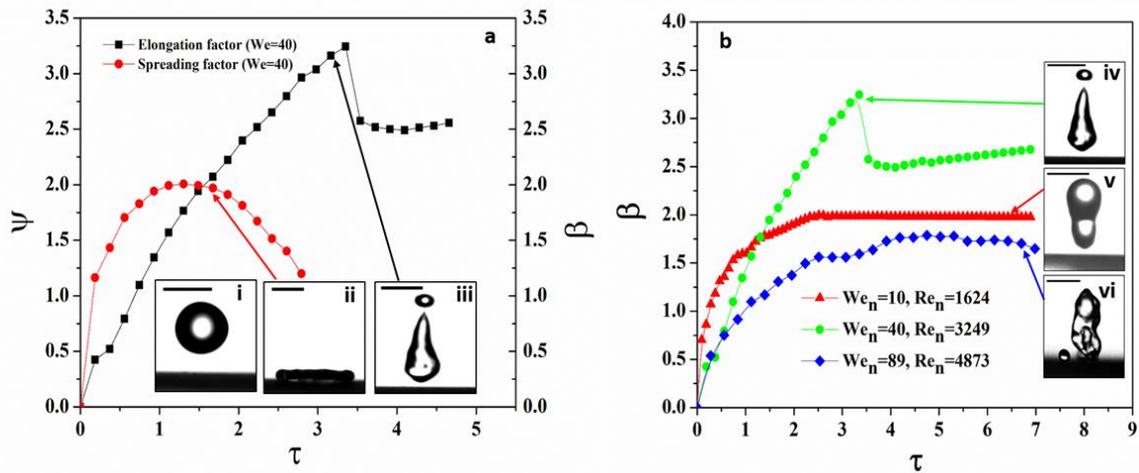

**Fig.8** (a) Variation of elongation factor (β) and spreading factor (Ψ) with non-dimensional time on SH surface for We= 40 on horizontal surfaces. Left hand ordinate represents β and right hand represents Ψ. Inset: (i)-(iii) show the initial drop with scale bar equal to 2.9mm, maximum spreading diameter at 3ms, and the separation of secondary drop at 10.26ms, respectively. (b) Elongation factor vs. non-dimensional time at different We. Inset (iv)-(vi) shows elongation dynamics of water drops post impact on horizontal SH surface.

From Fig.8 (b) it is evident that the temporal variation of elongation factor on horizontal SH surfaces is strongly dependent on Weber number. The inset shown in Fig.8 (b) corresponds to the drop structures for each We in their final state of the present measurement time span (t= 10.26 ms). From the captured time-series images it has been observed that the recoil causes the drops to elongate vertically and aids the formation of secondary drops at the initiation of recoiling. From the inset of Fig. 8b-(v) it can be observed that at lower $We_n$ =10 no secondary drops are formed. At $We_n$=40 elongation factor increases initially followed by a sudden drop due to formation of a secondary droplet. At We=89, elongation factor is suppressed due to breaking of the rim into secondary drops at the time of maximum spreading.

Further, the effect of surface inclination on the temporal evolution of elongation factor β at a fixed We=40 ($We_n$ is changing according to the inclination) has been shown in Fig.9. It has been observed that the formation of secondary drop is inhibited with the increase of surface inclination. If the sum of inertial and surface energies during upward acceleration exceeds the initial surface energy during recoiling of droplet, the bouncing and pinching off behavior of post impact droplet can be observed[40, 43]. Hence the inertial energy along the normal direction stimulates the pinching off behavior to produce secondary drop. Reduction in normal velocity component ($v_i \cos\Phi$) at inclined conditions decreases the temporal variation of elongation factor and suppresses the pinch-off. In addition, the tangential velocity component ($v_i \sin\Phi$) supports the post impact drops to slide along the inclined surface, thereby diverting a fraction of the kinetic energy which otherwise promotes orthogonal pinch-off.

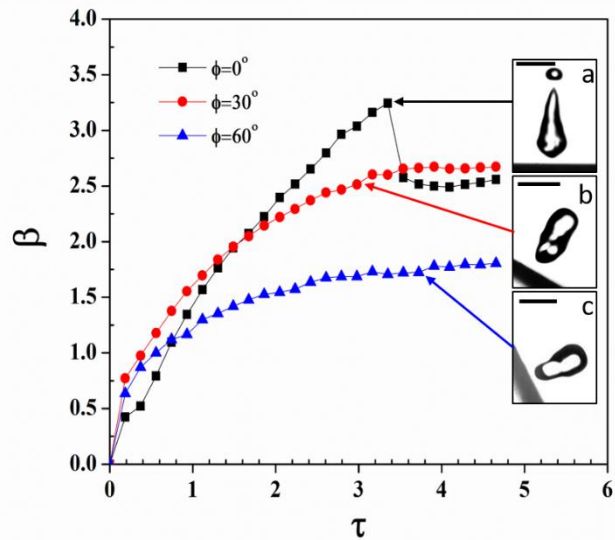

**Fig.9** Effect of surface inclination on elongation factor for water droplet on SH surface. Inset: (a)-(c) Images of post impact drop on the inclined surface at t=10.26ms. The scale represents 2.9 mm for each image. Based on initial droplet velocity before impact, the conditions are of We=40 and Re=3249. At 30° and 60°, the $We_n$ are equal to 30 and 10 respectively; similarly, normal Reynolds number ($Re_n$) are equal to 2814 and 1625 respectively.

### III.     Effect of surface inclination on energy ratio

Fig.10 illustrates the variation of pre and post impact non-dimensional kinetic energy ($E^{"}_{kine,normal} = We_n/3$), adhesion energy ($E^{"}_a = \left[(1-\cos\theta)\psi^2_m - (l_m/r_i)\psi_m \sin\theta\right]$), surface energy ( $E_d^{"}=S_d/\pi r_i^2$), non-dimensional gravitational potential [$E_{grav}^{"}=2\rho l g r_0 h_m/(3\sigma)$] and viscous dissipation $\left[ E_v^{"}=\frac{27}{16}\frac{r_i^2}{l_m^2}\frac{We}{Re}\psi_m \right]$ components at a fixed We=10. Each energy component has been non-dimensionalized by the initial surface energy ($E_{sprd}=\sigma\pi d_i^2$). It is evident that the energy components like surface deformation energy, adhesion and kinetic energy in normal direction decrease with increase in inclination angle, thereby resulting in decrease of the maximum spreading factor with the increase of inclination. Reduction in adhesion energy is manifested as faster movement of post impact droplet along the plane.

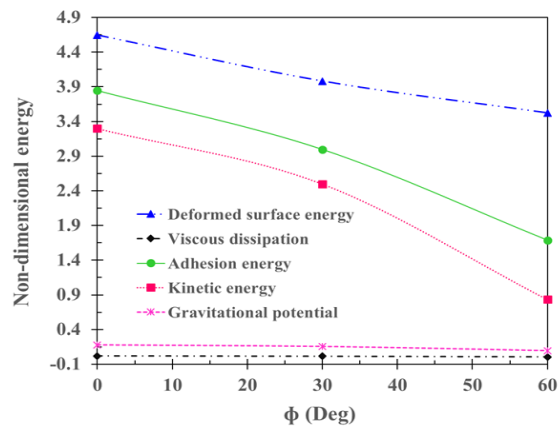

**Fig.10** Variation of components of the non-dimensional energy with surface inclination, for SH surface and We=10.

Fig.11 illustrates the effect of surface inclination and Weber number on energy ratio ($E_r$), non-dimensional viscous dissipation ($E_v^{"}$), and adhesion energy ($E_a^{"}$) respectively, for SH surfaces. The energy ratio has been defined as the ratio of non-dimensional viscous dissipation to the adhesion energy. From Fig.11 it can be observed that $E_r$ decreases with increasing inclination angle for a fixed We. At lower We (=10), the $E_r$ stays fairly constant. This is because both the non-dimensional adhesion ($E_a^{"}$) and viscous ($E_v^{"}$) energy components are dependent on the inclination angle, as shown in Fig. 10 (b). Therefore the decrease in maximum spreading factor along a tilted surface can be attributed to the reduction in the adhesion and deformed surface energy components.

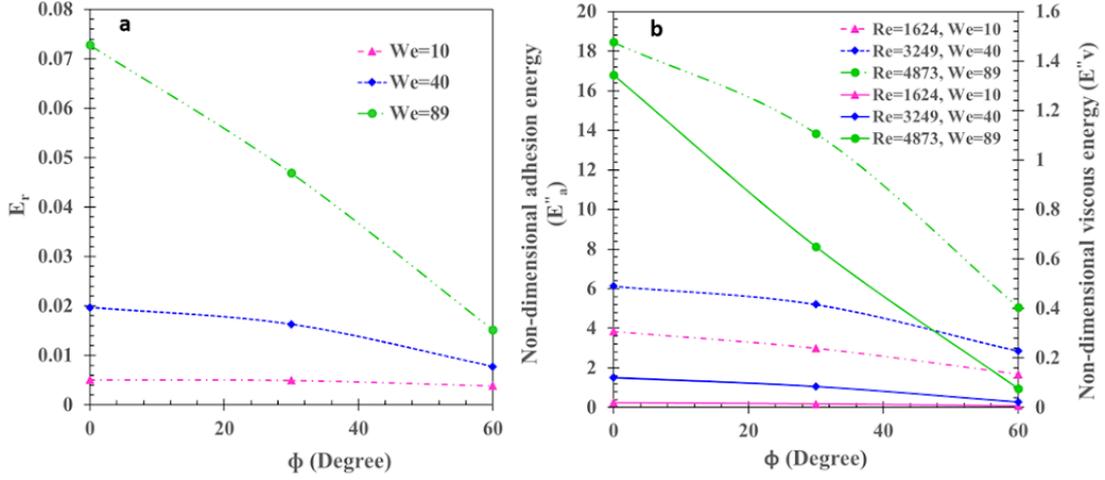

**Fig. 11** (a) Variation of energy ratio vs surface inclination (b) Variation of non-dimensional energy vs surface inclination for SH surfaces. The left hand side ordinate (dotted lines in the plot) shows the non-dimensional adhesion energy. Similarly, the solid lines of the same color as the dotted ones represent the non-dimensional viscous dissipation for same Re and We in right hand side ordinate.

**b. Hydrophilic surface**

**(i) Maximum spreading factor ($\Psi_m$) and maximum spreading time $\tau_m^*$**

Analogous to the SH surface investigations, initial focus was on determination of maximum spreading factor ($\Psi_m$) and maximum spreading time ($\tau_m^*$) of droplets impacting on hydrophilic surfaces. In order to investigate the spreading dynamics along the inclined hydrophilic surfaces, the average contact angle ($\theta_{av}$ in eqn.6) has been used in eqn. 1. The dynamic contact angles measured at the instant of maximum spreading diameter have been tabulated in Table S2 (supporting data). The analytical expression (eqn. 1) involving spreading factor $\Psi_m$ has a quadratic nature. Of the two roots of eqn. 1, the root of lower value accurately approaches the experimental results (explained in supplementary information). The deformed surface $S_d$ to be used in eqn. 1 considers two separate expressions (eqns. 5 & 7 depending on the inclination angle).

From Fig. 12, it is observed that the maximum spreading factor increases with an increasing inclination angle. This is in contrast to SH surfaces (fig. 5) where decreasing trend of maximum spreading factor vs. inclination angle at a fixed We is noticed. The modified theoretical model agrees to an extent with the experimental results at low inclination angles (fig.12 a). However, it is observed that the analytical result diverges from experimental observations beyond Φ>40°. At higher inclinations, pinching off produces secondary droplets at the corner point[41] (inset iii of Fig. 12b) leading to deviations from the analytical predictions. Fig.12(b) illustrates the temporal evolution of spreading factor $\Psi_m$ at different inclination angles of hydrophilic surfaces. The values of spreading factor (ψ) after non-dimensional time τ =2 tends to attain a saturated value. The saturated values of ψ are observed to enhance with increase in the inclination angle. As the tangential velocity component $v_i\sin\Phi$ increases with inclination Φ, drops spread more along the hydrophilic surface resulting in increasing trend of maximum spreading factor vs. inclination angle.

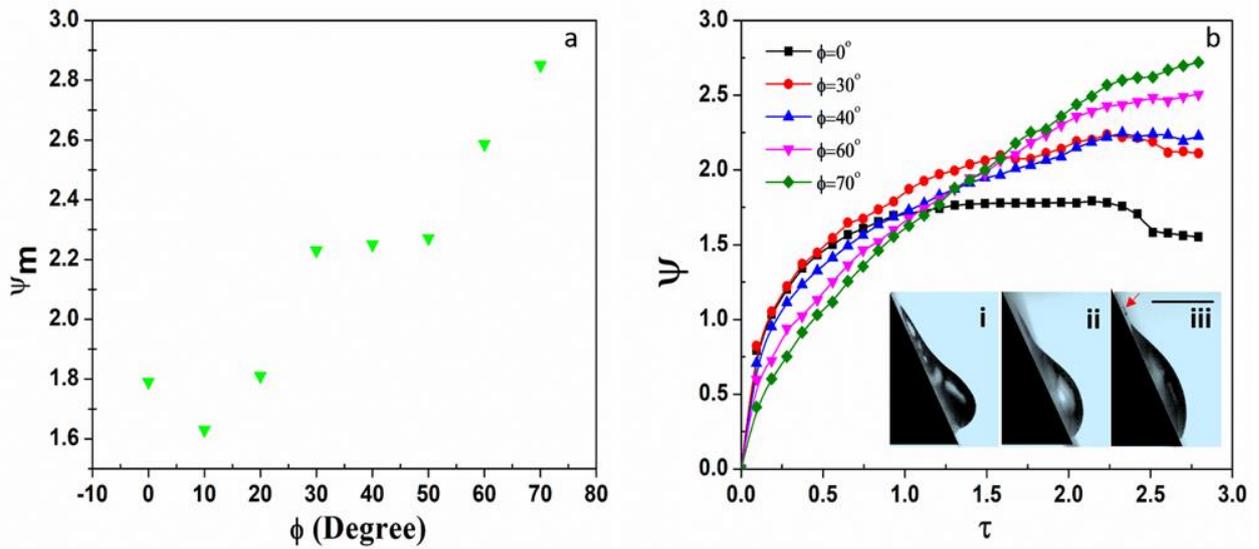

**Fig.12** (a) Maximum spreading factor vs. surface inclination on hydrophilic surface with impact velocity of 0.5 m/s for water droplets (We=10, $We_n = We*(\cos(\Phi))^2$ ). (b) Temporal variation of spreading factor for different Φ for hydrophilic surface with impact velocity of 0.5 m/s. Inset: (i), (ii) and (iii) show the side view of spreading of post impact water drops at Φ=60°. The images were taken at t=14.85, 27, and 50.27 ms, respectively. The red arrow in (iii) illustrates the formation of secondary drops due to pinching off phenomenon. The scale in part b inset (iii) corresponds to 2.9 mm.

Figure 13 a illustrates the normalized maximum spreading time ($\tau_m^*$) with Weber number for a horizontal surface ($We_n$=We for Φ=0°). Figure13a is analogous to horizontal SH surfaces (figure8a-d blue triangular symbols denoting $We_n$=We for Φ=0°). It is evident that spreading time decreases with Weber number with a power law scaling of -0.6153. With the change of inclination angle Φ, normal Weber number $We_n$=WecosΦ decreases. Since maximum spreading factor is increasing with inclination angle (fig.13a), spreading time is supposed to show increasing trend. However, in contrast to SH surfaces (fig.8), figure 13b shows the reverse trend i.e. the spreading time is increasing with the increase in $We_n$ (or decreasing with increasing angle Φ). So although for horizontal hydrophilic surfaces eqn.8 can be suitably modified with change in prefactor and scaling exponent, inclined hydrophilic surfaces can't be described by eqn. 8 as it uses a negative scaling exponent. This indicates the dependence of spreading time on some other governing parameter besides $We_n$, which will be explored in fig. 14b.

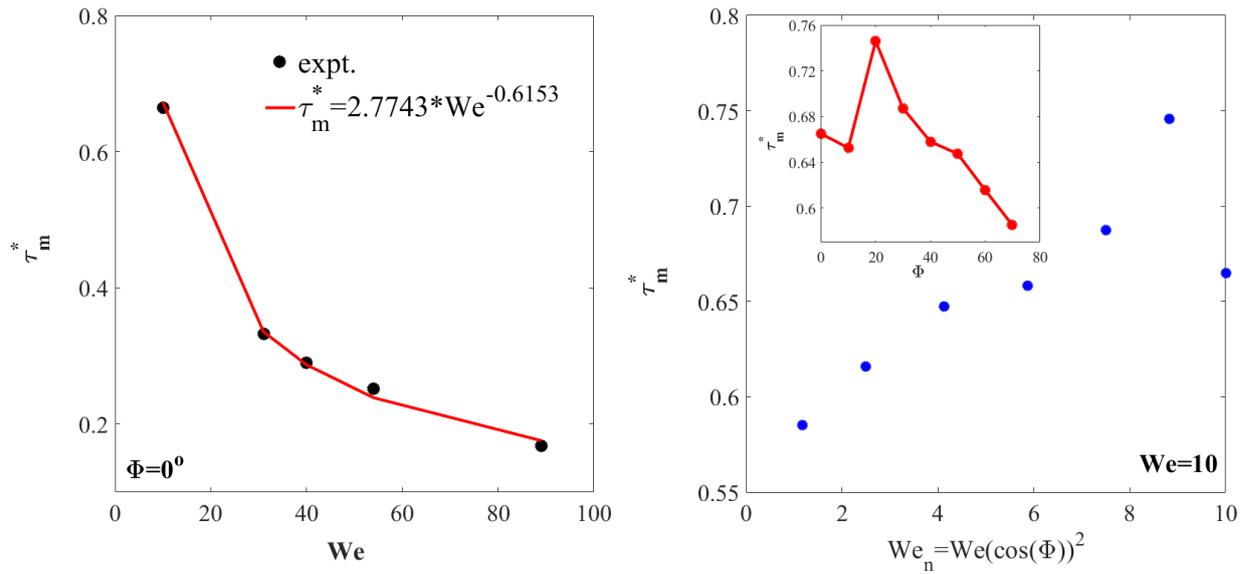

**Fig.13 (a)** Variation of Normalized spreading time with We for horizontal hydrophilic surface (Φ=0°) **(b)** Normalized spreading time vs. $We_n$ with changing surface inclination at impact velocity of 0.5m/s on inclined hydrophilic surfaces with water droplets (We=10 corresponds to horizontal surface, $We_n = We(\cos(\Phi))^2$ decreases with increasing angle Φ for a constant We) . Inset shows the dependence of Normalized spreading time vs. inclination angle.

**(ii)   Effect of surface inclination on energy ratio**

Fig.14 (a) illustrates the energy components for hydrophilic impacts. While the non dimensional viscous energy is of similar magnitude with the SH surfaces, non dimensional adhesion energy is one order lesser compared to SH surface (fig. 11b and figure 14a for comparison). Since the energy ratio $E_r$ is defined as the ratio of nondimensional viscous energy to non dimensional adhesion energy, $E_r$ significantly decreases up to a certain inclination angle (~ 35–40°) and then remains nearly constant with the surface inclination (fig. 14a inset).

For hydrophilic surfaces on inclined planes, sliding occurs. In case of sliding droplets, tangential kinetic energy is calculated instead of normal kinetic energy. As seen from 14b, the magnitude of non-dimensional tangential kinetic energy ($E''_{kine, tangential}=We_t/3$) is dominant over non-dimensional adhesion energy $E''_a$ and the non-dimensional gravitational potential along the inclined hydrophilic surfaces. Thereby it is inferred that the tangential component is the critical parameter to promote spreading process of post impact drops along inclined hydrophilic surfaces. So in spite of the increase of maximum spreading factor with increasing $\Phi$ (fig.12), increase of tangential kinetic energy with inclination angle is leading to lower $\tau_m^*$ at higher inclination angles or lower $We_n$ (fig.13b).

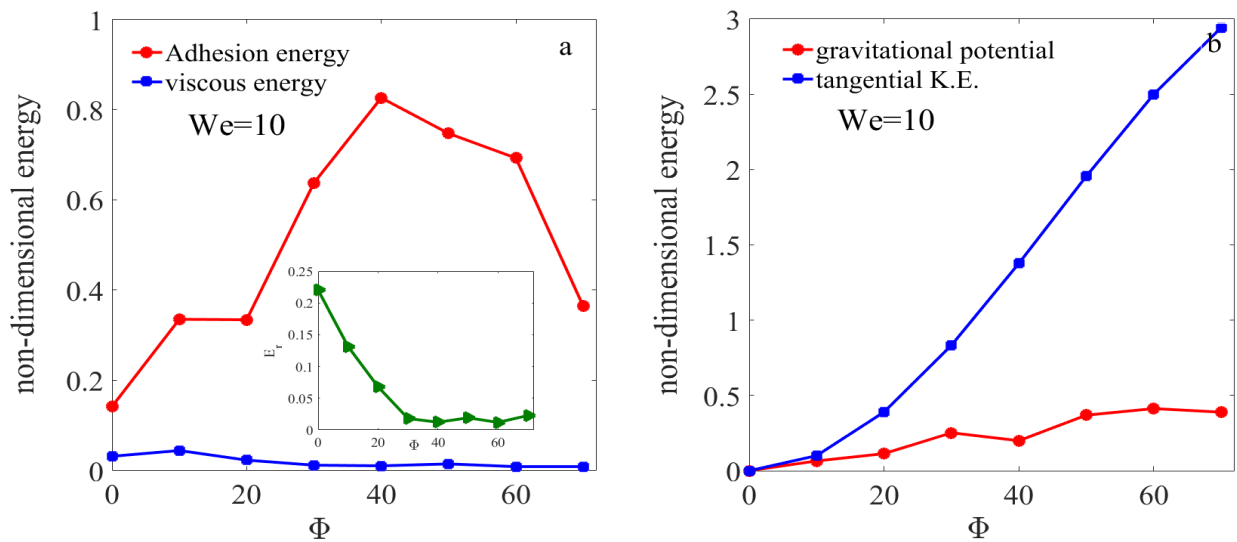

**Fig. 14** (a) Effect of surface inclination on the energy ratio ($E_r$), the non-dimensional viscous dissipation ($E_v$), and the adhesion energy ($E_a$), on hydrophilic surfaces; for water droplets. Inset shows the energy ratio ($E_r=E_v/E_a$) vs. $\Phi$ (b) Variation of non-dimensional gravitational potential (right hand ordinates) and tangential kinetic energy (left hand ordinates) vs. surface inclination. We=10 if angle variations are not considered.

**(c) Nanocolloids as a special fluid to reveal the role of slip and disjoining pressure on SH and hydrophilic surfaces respectively**

The mechanisms governing spreading dynamics of post impact drops in SH and hydrophilic surfaces are grossly different (save for the common influential parameters). In this section, the role of slip and disjoining pressures on surfaces of varying wettability will be discussed using nanocolloids as the test fluid. The slip phenomenon at the nanoscale plays an important role in case of SH surfaces and dispersed particles in base fluid alters the slip length on SH surfaces[39]. However, in hydrophilic surfaces, the disjoining pressure dictates the spreading performance of impacting colloidal drops[44]. Accordingly, the mechanisms for impact of the colloidal drops need to be discussed. Spreading of fluids on surfaces is characterized by the interfacial tensions and the film energy, which is expressed by the integral over the film thickness of the disjoining pressure. The disjoining pressure is expressed as a combination of the van der Waals, electrostatic, and the structural components. In case of colloids, the structural component of disjoining pressure dominates over the other two [45]. The disjoining pressure is the difference between the pressure in a region of a phase adjacent to a surface confining it and the pressure in the bulk of this phase[46]. An analytical expression for structural component of the disjoining pressure can be expressed as [47]

$$\prod_{st}(h) = \prod_1 \cos(\omega h + \varphi_2) * e^{kh} + \prod_2 * e^{-\delta(h-d)} \text{ for } h \geq d \tag{9}$$

$$\prod_{st}(h) = -P \text{ for } 0 \leq h < d \tag{10}$$

where d is the nanoparticle diameter; h is the microfilm thickness formed by wetting; δ and ω are the decaying parameter and all other parameters considered in Eqn. 9 are cubic polynomial fitting parameters ($\Pi_1, \varphi_2, k$) based on volume fraction of the colloid. It is related to the particle volume fraction as $\Phi = \left(\dfrac{6n_p}{\pi d^3}\right)$. Here, $n_p$ is the number of particle per unit volume and P is the bulk osmotic pressure.

Due to the formation of electrostatic double layer (EDL), the effective particle diameter needs to be considered as the summation of particle diameter and double layer thickness. As the particle concentration increases, it results in the enhancement of structural component of disjoining pressure which enhances the spreading of colloids[48]. This is evident in Fig. 15 (b), where, despite largely augmented viscosity, the silica colloids spread to larger extents than water on hydrophilic surfaces. The slip length is a direct manifestation of deviation from the no-slip mode at the wall to the Navier slip condition during spreading of liquid along the interface. The magnitude of slip length determines the interactivity of the nanoparticles at the fluid–surface interface and their influence on the nature of motion of the colloidal contact line. Here, the classical Cox-Voinov-Tanner law is invoked and the slip length is deduced scaled from information of the receding contact angle of different fluids.

The difference between the cubes of the static and dynamic receding contact angles is expressed as a function of the contact line capillary number and a power law formulation is framed[49, 50]. It is observed that the nanocolloidal drops change the magnitude of slip length compared to that of the base fluid. The index of the contact line capillary number for a given fluid on a given surface indicates the magnitude of the deviatory behavior of the dynamic advancing angle with respect to the static angle. This thereby provides information regarding the propensity of the fluid to wet or de-wet the particular surface[51].

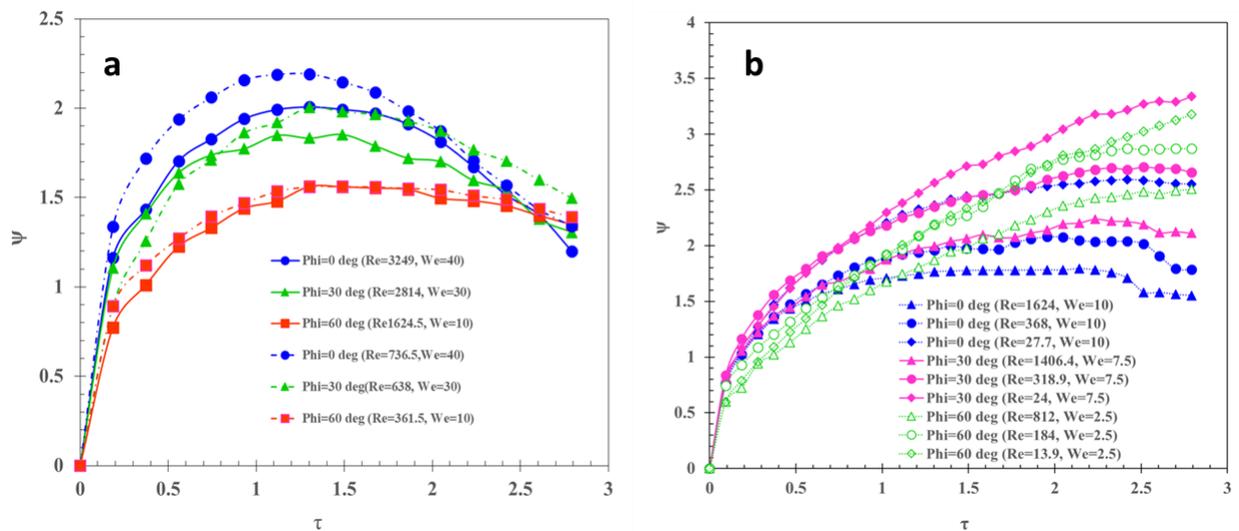

**Fig.15** (a) Temporal evolution of spreading factor on SH surfaces at We=40. The liquids are water and 2.5 wt % silica nanocolloid, (b) temporal evolution of spreading factor on hydrophilic surface at We=10. The liquids are water, 2.5 and 5 wt % silica nanocolloids.

Fig.15 illustrates the temporal evolution of spreading factor on SH and hydrophilic surfaces for water and silica nanocolloids. The nanocolloids show enhanced spreading factor marginally at lower or zero inclination angle as illustrated in Fig.15 (a). At higher inclination, gravitational effect becomes dominant and enhances inertia along the surface. In the case of horizontal surface spreading, inertia force and disjoining pressure aids the spreading whereas surface tension and viscous dissipation acts against it. In general, it is reported[52] that as the concentration of particle increases, there is only a marginal increase in surface tension of the nanofluids. Hence, the role of surface tension force is negligible in determining the spreading behavior of colloids. When a colloidal drop approaches a smooth hydrophilic solid surface, there is a microscopic transition between the liquid film and the meniscus and has a wedge-like profile. Its shape is determined by forces arising from the ordering of particles[45]. As the film thickness decreases towards the wedge vertex, the structural disjoining pressure increases. The structural component of disjoining pressure corresponding to a film thickness with one layer of particle is observed to be higher than that with two or more layers of particle[53]. Consequently there is a film tension gradient generated towards the wedge from the bulk of solution with a higher film tension near the vertex. As the film tension increases towards the vertex of wedge, it aids the nanocolloid to spread further. On increasing the inclination angle, the inertial component is assisted by the gravitational potential which acts as the deciding factor for spreading behavior.

## 5. CONCLUSIONS

The present study focused on the impact dynamics of drops on surfaces of different wettability and different inclinations. The principal observations and findings can be summarized briefly as:

- The maximum spreading factor and time for SH surfaces is found to decrease with inclination angle whereas hydrophilic surfaces shows the reveres trend. The corresponding trends can be explained from their respective energy components. In case of SH surfaces, all the energy components are decreasing with increasing inclination angle. In case of hydrophilic surfaces where droplets slide after impact, tangential kinetic energy needs to be taken into account. Non-dimensional tangential kinetic energy is shown to be dominant over adhesion energy and gravitational potential and increasing with inclination angle. This explains the decreasing trend of spreading time with inclination angle in spite of increasing nature of spreading factor with increasing angle. The modified Y-K model with $We_n$ and $Re_n$ was in good agreement with SH surfaces whereas it was not satisfactory for the hydrophilic surfaces.
- In addition, elongation factor of droplets, coefficient of restitution on impact upon SH surfaces has been calculated. Dependence of viscosity on spreading factor is also studied. Arrest of secondary droplet pinching off at higher inclination angle of SH surfaces was observed.
- For nanocolloids, the slip phenomenon characterizes the spreading behavior of droplets on SH surfaces whereas the disjoining pressure is the critical parameter for enhancement of spreading in hydrophilic surfaces.

Based on the above inferences, the present findings will be beneficial in controlled deposition of drops on generic inclined surfaces. The present findings could have potential applications in better design and operation of droplet deposition and spray technologies.

## Acknowledgements

NS, GK and ARH would like to thank the Ministry of Human Resource Development, Govt. of India, for the doctoral scholarships. DS and PD would like to thank IIT Ropar for funding the present work (vide grants 9-246/2016/IITRPR/144 & IITRPR/Research/193 respectively).